\documentclass[aps,prd,preprintnumbers]{revtex4-1}

\usepackage{graphicx}
\usepackage{color}
\usepackage{slashed}
\usepackage{amsmath}
\usepackage{amssymb}

\newcommand\RNum[1]{\uppercase\expandafter{\romannumeral #1\relax}}
\newcommand\bef{\begin{figure}}
\newcommand\eef[1]{\label{fg:#1}\end{figure}}
\newcommand\beq{\begin{equation}}
\newcommand\eeq[1]{\label{#1}\end{equation}}
\newcommand\beqa{\begin{eqnarray}}
\newcommand\eeqa[1]{\label{#1}\end{eqnarray}}
\newcommand\bet{\begin{table}}
\newcommand\eet[1]{\label{tb:#1}\end{table}}
\newcommand\fgn[1]{Figure \ref{fg:#1}}
\newcommand\eqn[1]{eq.\ (\ref{#1})}
\newcommand\eqand[2]{eqs.\ (\ref{#1}) and (\ref{#2})}

\newcommand\apn[1]{Appendix \ref{sec:#1}}

\newcommand\ie{{\sl i.e.\/}}

\newcommand{\alphas}{\alpha_{\scriptscriptstyle S}}
\newcommand{\bilin}[1]{\overline Q{#1}Q}
\newcommand{\bsigma}{\boldsymbol\sigma}
\newcommand{\CM}{{{\scriptscriptstyle CM}}}
\newcommand{\EM}{{{\scriptscriptstyle EM}}}
\newcommand{\heq}{\overset{\scriptscriptstyle HQRF}{=\joinrel=\joinrel=}}
\newcommand{\lqcd}{\Lambda_{\overline{\scriptscriptstyle MS}}}
\newcommand{\M}{{\cal M}}
\newcommand\Nabla{\boldsymbol\nabla}
\newcommand\OLambda{\overline\Lambda}
\newcommand\PP{{\cal P}}
\newcommand\PPi{{\widetilde\Pi}}
\newcommand{\pt}{p_{\scriptscriptstyle T}}
\newcommand{\raat}{R_{AA}}
\newcommand{\tco}{T_{co}}
\newcommand{\Vc}[1]{{\mathbf #1}}

\begin{document}

\title{Spin polarization of heavy quarks in matter:\\ predictions from effective field theories}
\author{Sourendu Gupta}
\email{sgupta@theory.tifr.res.in}
\affiliation{Department of Theoretical Physics, Tata Institute of Fundamental
         Research,\\ Homi Bhabha Road, Mumbai 400005, India.}
\begin{abstract}
The spin polarization of heavy quarks in heavy-ion collisions at the
LHC is estimated from effective field theories (EFTs). One EFT is
similar to the HQET used at zero temperature. This gives a coupling of
the heavy quark spin to colour and electromagnetic fields in heavy-ion
collisions. The second EFT describes the interaction of heavy quarks and
hydrodynamic modes, and gives the coupling between the heavy quark spin
and the local vorticity of the fireball. Using these, we find that the
measurement of polarization of the heavy quark from small to moderate
$\pt$ at the LHC is predicted with a single free parameter proportional
to the vorticity. As a result, the heavy quark polarization is the same
whether it is derived from the spin alignment of heavy vector mesons or
the polarization of heavy baryons. We also predict that the parameter does
not differ much between charm and bottom quarks.
\end{abstract} \maketitle

\section{Introduction}

Perhaps one of the most interesting recent observations in the study of
heavy-ion physics has been the spin polarization of hadrons. Until now
there have been measurements of the polarization of $\Lambda$, $\Xi$, and
$\Omega$ baryons \cite{STAR:2017ckg, ALICE:2019onw, STAR:2020xbm} as well
as spin alignment of $\rho$ and $K^*$ vector mesons \cite{ALICE:2019aid,
STAR:2022fan}. The spin polarization (or alignment) has to be due to
an ordering axial vector field, which could be electromagnetic fields
produced by the two charged nuclei travelling parallel to each other
with relativistic speeds, or the overall angular momentum of the fireball.

Parallely, a substantial improvement has come about in the understanding
of the interaction of heavy quarks with thermal QCD matter, largely by
exploiting the hierarchy of scale between the heavy quark mass, $m$,
and the temperature, $T$, of matter. On the theoretical side, there has
been steady progress over time as details of heavy quark thermalization,
and the relevant transport coefficients, were worked out in studies using
weak coupling theory \cite{Braaten:1991we, Moore:2004tg}, the lattice
\cite{Banerjee:2011ra, Ding:2012sp, Altenkort:2020fgs}, and effective
field theories (EFTs) \cite{He:2011qa, Cao:2013ita}. Experiments at
the LHC have studied $\raat$, the ratio of cross sections in AA and
pp collisions as functions of transverse momentum, $\pt$, and impact
parameter at central rapidity, \ie, $y<0.5$--1, for multiple nuclei. It
turns out that both $\raat$ and the azimuthal flow, $v_2$, of charm
can be described in transport theory, and yield values for the charm
quark momentum diffusion constant \cite{ALICE:2020iug, ALICE:2021rxa}
in agreement with lattice results.  The momentum diffusion constant
yields charm thermalization time in the range of 3--8 fm. Since this
time scales as the heavy quark mass, the bottom thermalization time is
expected to be about 10 fm or more. As a result, bottom quarks are not
expected to be thermalized. Similarly, the momentum of charm quarks in
non-central collisions at the LHC, and for high-$\pt$ even in
central collisions are also at best partially thermalized.

\bef[bt]
\begin{center}
	\includegraphics[scale=0.6]{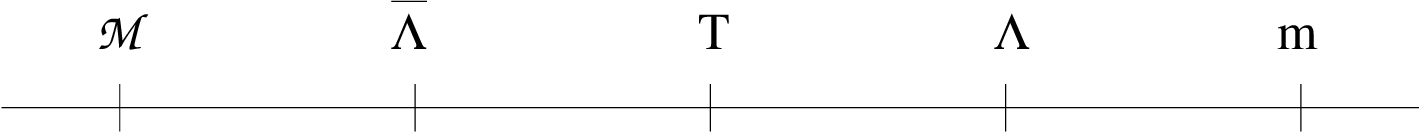}
\end{center}
\caption{The major energy scales considered here are (lowest on the left)
 the inverse of the typical size or lifetime of the fireball, $\M$,
 the temperature of the medium, $T$, and the heavy quark mass,
 $m$.  Thermal HQET is defined with a UV cutoff $\Lambda$ such that
 $m>\Lambda\gg T$. HQSD is defined with a UV cutoff $\OLambda$ satisfying
 $T>\OLambda\gg\M$, so that the medium can be described by a colour
 neutral fluid.}
\eef{scales}

Here we suggest that spin polarization experiments with high $\pt$
heavy flavour hadrons, especially of bottom flavour, are of significant
interest. The measurement is interesting in itself \cite{ALICE:2022pxx}.
Moreover, by writing two bottom-up EFTs for the interaction of a
heavy quark with matter, we demonstrate here that the measurement
of polarization of heavy flavour hadrons would give theoretically
controllable insights. EFTs are written as usual by choosing an
appropriate cutoff, and using the symmetries of the problem to write
all possible terms in the Lagrangian organized by mass dimension.

The light degrees of freedom in the fireball, gluons and light quarks,
are characterised by the energy and momentum scale $T$. In the early
stages of the fireball one has $T>\tco$, where $\tco\approx\lqcd$.
Realistically, $T$ may be as large at 0.5 GeV at early times, but not
much higher.  One could take $m=4.18\pm0.03$ GeV for bottom quarks
and $m=1.280\pm0.025$ GeV for charm \cite{ParticleDataGroup:2020ssz}.
By choosing a cutoff $m>\Lambda\gg T$, one may write an EFT for the heavy
quark which is not very different from the vacuum heavy quark effective
theory (HQET). Clearly, the EFT is more likely to be under control for
the bottom quark than the charm. Vacuum HQET finds quantitative use in
computing the decays of hadrons with heavy quarks. For $T>\tco$ there
are no hadrons and HQET will be useful in finding the polarization of
heavy hadrons.  This polarization would carry through the hadronization,
since heavy quark spin symmetry implies that dynamics at longer scales
does not change it.

At distance larger than the transport length scale (in a weakly coupled
medium this is the mean free path), matter can be described as a fluid. If
$g(T)$, the strong coupling at the scale $T$, is small enough, then a
power counting shows that the transport scale $1/[Tg^4(T)\log g(T)]$ is
longer than the magnetic screening length $1/[Tg^2(T)]$, so the fluid
would be colourless. However, the approximation $g(T)\ll1$ cannot be
used, as becomes clear when we consider a dimensionless fluid variable
called the liquidity, $\ell$, which is the ratio of the transport length
scale to the typical microscopic distance scale \cite{Gupta:2003be}. In
the weak coupling regime $\ell=1/[g^4(T)\log g(T)]\gg1$. In this
regime the fluid behaves like a gas, which has mean free path much
larger than the interparticle spacing. However, one can also write
$\ell\approx\sqrt[3]{S/\eta}$. In the fireball one finds $\ell\approx1$,
which means the fluid behaves like a liquid, so $g(T)$ is not small,
and the weak-coupling expansion is not applicable at energy scale of
order $T$. By choosing $\Lambda\gg T$ in HQET we ensure that the weak
coupling expansion is not pushed to a scale where it is not applicable.

However, experimental studies of heavy-ion collisions have a good
description in terms of a colourless fluid \cite{Bernhard:2016tnd,
JETSCAPE:2020mzn}. So at length scales where this fluid description works,
one should use variables which describe fluid flow and its fluctuations
\ie, phonons, as well as flavour current. Since QCD does not allow
flavour mixing, the flavour current of the heavy quark has no direct
current-current coupling to other flavour currents; at hydrodynamic scales the
currents are only coupled via phonons, as a result of which the couplings
appear at mass dimension greater than 6. Therefore, the leading couplings
of a heavy quark current are to flow vectors and phonons.  The momenta
of phonons would be bounded, of course, with $|\Vc k|<\OLambda$.  Also,
since the spin-symmetry of the heavy quark persists into the IR, a compact
description of the heavy-flavour current and its spin is provided by a
colour neutral heavy fermion field. So the hydrodynamic effective theory
below the UV cutoff $\OLambda$ may be called Heavy Quark Sono-Dynamics
(HQSD).  Take the scale of the fireball (typical spatial size or lifetime)
to be $1/\M$. Since the Knudsen number $\kappa=\M/T\ll1$, one
may use a UV cutoff $\OLambda$ for HQSD, with $T>\OLambda\gg
\M$.

In the Sections \RNum2 and \RNum3 we discuss the Lagrangians of HQET
in matter and HQSD respectively. Applications to heavy quark spin
thermalization and polarization in heavy-ion collisions at the LHC
can be found in Section \RNum4. Our conclusions and a summary of the
observations which could give evidence of the physics discussed in this
paper are in Section \RNum5. Technical material, such as details of the
objects needed for the construction of EFTs, appears in the appendix.

\section{Heavy Quark Effective Theory in matter}

In vacuum HQET \cite{Isgur:1989vq, Georgi:1991mr, Neubert:1996wg} light
degrees of freedom are integrated out to some scale $\Lambda$ such that
$m>\Lambda\gg\lqcd$. The second inequality ensures that this can be done
in weak coupling. The operator product expansion required for writing the
EFT does not change in thermal matter, although the tensor composition
of the local operators does (see Appendix A for details). Further, when
$\Lambda\gg T$, the weak coupling computations are also very similar. As
a result, HQET in matter does not differ too much from the more familiar
case of HQET in vacuum.

HQET in matter has two consequences. The first is heavy-quark velocity
superselection, similar to that in the vacuum theory.  The momentum
of a heavy quark can be written in terms of its 4-velocity, $v$, in
the form $p=mv+k$ where the momentum $k$ comes from the soft kicks
given to the heavy quark by the gluons in the medium. By demanding
that $v\gg\Lambda/m$, one controls these kicks when the transverse
momentum of the heavy quark is larger than the cutoff, $\pt\gg\Lambda$.
Velocity superselection allows us to treat the heavy quark rest frame
(HQRF) as an inertial frame and write Lorentz invariant quantities more
simply in the HQRF (see Appendix A).  The second is that choosing
$\Lambda\gg T$ allows the integration of hard modes to be done in weak
coupling theory. This avoids the problem of having to use $g(T)$, which,
as we argued before is large. The choice of UV cutoff then allows us to
match the LECs using weak coupling theory, at least for bottom.

For $v\gg\Lambda/m$ one first factors the conserved quantity $v$
by defining the Dirac field $\chi = \exp(imv\cdot x)\psi$. Since the
remainder of the heavy quark momentum is smaller than $\Lambda$, it is
useful to decompose $\chi$ into large and small parts, the large part
being $Q=P_+\chi$ where $P_+=(1+\slashed v)/2$. Then using the methods
of \apn{frames}, one finds that in the HQRF one can write
\beq
 L_{HQET}
   = \frac i2 \bilin{\partial_t}
     - \frac{c_1}{2m} \bilin{{\mathbf D}^2} 
     + \frac{c_2}{4m} \bilin{\Sigma_{\mu\nu}}gF^{\mu\nu}
     + {\cal O}\left(\frac1{m^2}\right)
\eeq{hqet} 
where $\partial_t$ is a time derivative, $\mathbf D$ is the spatial
part of the covariant derivative, $g$ is the strong coupling taken
at the scale $\Lambda$, and $F^{\mu\nu}$ is the field strength tensor
for gluons.  The LECs $c_{1,2} = 1 + {\cal O}(\alphas)$ are computable
in the weak-coupling expansion.  Loop corrections to the heavy quark
propagator can induce a mass correction of the form $\Delta m =
{\cal O}(\alphas)$. The form of \eqn{hqet} is robust; when $g\simeq1$
it is the ability to compute the LECs in weak-coupling which is lost
\cite{Bouttefeux:2020ycy}.  Note that the piece of the action with the
spin operator, $\Sigma_{\mu\nu}$, commutes with the leading kinetic piece
of HQET, and gives rise to the emergent heavy quark spin symmetry. This
property is retained in the presence of matter.

The mass dimension 5 interaction term between the heavy quark spin and
the colour field is given by (see Appendix A),
\beq
 L_5 = \frac{c_2g}{2m}\times \begin{cases}
	 \bilin{\bsigma\cdot\Vc B}  & \rm{(in\ HQRF)} \\
	 \gamma_\CM\bilin{(\Vc v\cdot\Vc E+\bsigma\cdot\Vc B)}
	   & \rm{(in\ CM\ frame)} \\
       \end{cases}
\eeq{hamint}
There is no surprise about the form of the interaction, but HQET
gives the value of $c_2$ to whatever precision is required. There
is a source of large colour fields in the earliest stages of the
collision. At this epoch the momenta of the gluons is mostly in the $z$
direction, and a plasma mode called the Weibel instability is expected
to produce strong colour fields in the medium \cite{Mrowczynski:1993qm,
Romatschke:2003ms, Arnold:2003rq, Rebhan:2004ur, Bodeker:2007fw}.
Due to the Weibel instability, induced fields grow until non-Abelian
effects stop the growth. An estimate in \cite{Arnold:2003rq} is that
the saturation value for mode number $q$ is $gB\simeq q^2$ (see however,
\cite{Bodeker:2007fw}).  At temperature $T$ the most likely mode number
$q\simeq T$. So, by this argument, one would have $gB\simeq T^2$.
However, the direction of the colour magnetic field due to the Weibel
instability is fixed by initial perturbations, and will be different
from one event to another. Since polarization cannot be measured event
by event, unfortunately, this spin alignment cannot be used to yield
experimental evidence of the Weibel instability. A different use of the
effect of colour magnetic fields on the heavy quark spin is to estimate
whether or not the spin degree of freedom can be thermalized independently
of the momentum. We will present such an estimate in Section \RNum4.

The coupling of heavy quarks to electromagnetic (EM) fields is described
by simply adding the EM field tensor to \eqand{hqet}{hamint}. The EM
field has coupling $qe$ instead of $g$, where $q$ is the charge of the
heavy quark.  We do not add EM corrections to the LECs. This works since
$\alpha\simeq0.007$, and with $\alphas\simeq0.25$, the changes will be
numerically significant only at the level of three-loop strong-interaction
corrections.

In this discussion we have assumed that the cutoff $\Lambda$ can be
pushed to high enough values that weak coupling estimates are reliable.
As we discuss in Section \RNum4, this assumption may fail for the charm
quark, since $\Lambda$ has to be smaller than $m$. It is useful therefore
to note that a lattice computation of the LEC $c_2$ is feasible, but hard.
The spin polarization of the quark is clearly proportional to the magnetic
polarization. So from a linear response perspective, one may simply
define a magnetic susceptibility, $\chi$, by the constitutive relation,
$M=\chi B_\EM$, where $M$ is the magnetization of the heavy quark.
Due to the analogy between an imaginary chemical potential for a fermion
and an external U(1) magnetic field on the fermion, $\chi$ is related
to with the quark number susceptibility measured in lattice QCD, but
for a heavy quark. Introducing such a chemical potential for the heavy
quark, matching the lattice computation of the susceptibility to one
obtainable from \eqn{hamint} would give a non-perturbative matching of
$c_2$. This last step requires a knowledge of the charm $g-2$, relating
the magnetic moment and the spin. This is parametrically of the order
of $\alphas$ at the charm quark mass, and hence also requires a lattice
computation. This is the hardest part of the non-perturbative matching
of $c_2$ using the lattice.

\section{Heavy Quark Sono-Dynamics}

Next we turn to the extreme IR and the effective theory at momentum
scales smaller than typical transport time scales. This is the
regime of hydrodynamics. In this regime the heavy quark can only
interact with flavour currents, the conserved energy-momentum tensor,
and their fluctuations.  A colour bleached heavy flavour current is
easily constructed using (the large components of) a colour neutral
Dirac field.  QCD forbids any direct interaction between currents of
different flavours. Nor can such interactions be enabled by exchanging
hadronic excitations, since they all have effective masses of order
$T$, which lie beyond the cutoff $\OLambda$. However, there is always
one low-energy excitation in all hydrodynamic theories, namely sound
waves.  They are the Goldstone field of a broken conformal symmetry
and are described by a scalar field, since they have no polarization.
The theory of a heavy quark coupled to phonons is what we call Heavy
Quark Sono-Dynamics (HQSD).  We note that HQSD is equally valid for
charm and bottom quarks since $\OLambda\ll m$. In neither case can
the matching of LECs be done in weak coupling.

In order to control the construction of the theory, the effect of the
viscosity, $\eta$, has to be understood. Its major effect on sound is
to exponentially attenuate the waves in a typical length scale which is
called the Stokes length, $\ell_{\mathrm St}$. In a theory of phonons
this exponential decay may be modelled as an effective phonon mass,
$m_{\mathrm St} = 1/\ell_{\mathrm St}$. This mass scale can be written
in terms of $\eta$, the energy density of the fluid, $\epsilon$, and the
speed of sound, $c_s$, for a wave of frequency $\omega$ by the expression
\beq
 m_{\mathrm St} = \frac{2\eta\omega^2}{3\epsilon c_s^3}
   \approx \left(\frac\eta S\right)\;\frac{\OLambda^2}T.
\eeq{stokes}
The second expression is an order of magnitude estimate obtained by
setting $\epsilon\approx TS$, where $S$ is the entropy density of the
fluid, and taking the maximum possible value for $\omega\approx\OLambda$.
Then choosing $T=500$ MeV and $\OLambda=50$ MeV, as we argue in Section
\RNum4, we find that $m_{\mathrm St}$ is not larger than about 5 MeV. As
a result, the phonon effective mass due to viscosity can be neglected
within the lifetime of the fireball, and it may be treated as a true
Goldstone boson.

Since HQSD cannot be obtained directly from the QCD Lagrangian, we will
construct it bottom-up using the available symmetries. Taking into account
the existence of two special frames, namely HQRF and the fluid local
rest frame (FLRF), one can clearly write a Lorentz-invariant action
using the projectors given in \apn{frames}.  The simplest theory one
can write is given in the HQRF by
\beq
 L_{HQSD} = 
    +\frac12\left[\PPi_1^{\mu\nu}+c_s^2\PPi_2^{\mu\nu}\right]
       (\partial_\mu\phi)\,(\partial_\nu\phi)
    + \Delta m\bilin{} + \frac i2\bilin{\partial_t}
    - \frac{c_5}{2m}\bilin{{\Nabla}^2}
       + \cdots
\eeq{hqsd}
The first two terms are kinetic terms for the phonon (see Appendix A
for an explanation of why this term splits), the next term accounts
for all possible mass corrections in the theory, and the last two are
kinetic terms for the heavy quark.  All terms are of mass dimension 4,
except the mass term, which has dimension 3, and the last, which has mass
dimension 5.  As expected, there is no spin dependence here. Note also
that there are no interaction terms between $Q$ and $\phi$ at dimension
4 since terms such as $\bilin{}\,\phi$ are not allowed for the Goldstone
field $\phi$.

However, in setting up the Lagrangian in \eqn{hqsd}, we have
neglected the crucial fact that the fluid will generally have a local
vorticity pseudovector $\Vc w = \Nabla \times\Vc u$.  Since $\Vc u$ is
dimensionless, $\Vc w$ has mass dimension 1. Before proceeding, it is
useful to recall a formal property.  The vorticity $\Vc w=\Nabla\times\Vc
u$ gives rise to a topological invariant called the vortex helicity
\cite{Jackiw:2004nm}
\beq
  C = \int d^3x \Vc u\cdot\Vc w = \int d^3x \epsilon^{ijk}u_i\partial_ju_k.
\eeq{topo}
From the last expression it is clear that $C$ is similar to an Abelian
Chern-Simons term. The Kevin-Helmholtz theorem shows that $C$ is a
conserved quantity in a perfect fluid. More intuition for such a term
comes from electrodynamics where the 4-volume integral of the term
$\epsilon^{\mu\nu\lambda\rho}F_{\mu\nu}F_{\lambda\rho}$ can be reduced
to the form in \eqn{topo} with the vector potential taking the place of
$\Vc u$ and the magnetic field of $\Vc w$ \cite{Dunne:1998qy}.  What this
means is that a good generalization of the vorticity is an antisymmetric
rank-2 tensor, $\omega_{\mu\nu}$. The part of this which is analogous to
$\Vc B$ in $F_{\mu\nu}$ is $\Vc w$. The part that is analogous to $\Vc E$
we will denote $\boldsymbol\varpi$.

This generalization of the vorticity vector agrees with that used in
\cite{Becattini:2020ngo, Huang:2020dtn}, with the definition used
here being exactly a factor of two larger than that used in those
references. Furthermore, in the context of the polarization of hadrons
with light or strange quarks, two related quantities, the thermal
vorticity and the chiral vorticity, have been investigated. These lie
above the UV cutoff of HQSD, and so do not appear in the systematic
coarse graining used to construct this theory.

There is another critical issue to consider before constructing a theory
of heavy quarks coupled to the vorticity. For $\Vc w$ to be non-vanishing,
$\Vc u$ must change from one point to another. In that case the fluid
element coupled to the quark is accelerating. How is it then possible
to neglect technical issues connected to defining quantum field
theories in accelerated frames, such as the Unruh effect? One cannot,
in principle. Instead one can ask how important this effect is likely
to be in the current context. Since one expects $\Vc w$ to be similar
in magnitude to the expansion rate $\M$, both being macroscopic scales
in the fluid, the acceleration gives rise to an Unruh temperature of
magnitude $\M/(2\pi)\ll \OLambda < T$. So, in the thermal environment of
the fireball, one should be able to neglect the effect of acceleration
and use an EFT.

Since $\Vc w$ has mass dimension unity, it can lead to a new dimension 4
term in the Lagrangian,
\beq
 L_w = \frac{c_4}2\,\bilin{\Sigma^{\mu\nu}}\omega_{\mu\nu}
   = c_4 \times \begin{cases}
	   \bilin{\bsigma\cdot\Vc w} & \rm{(in\ HQRF)} \\
	   \gamma_\CM\bilin{(\Vc v\cdot\boldsymbol\varpi
	      +\bsigma\cdot\Vc w)} & \rm{(in\ CM\ frame)}\\
	 \end{cases}
\eeq{spinorbit}
where $c_4$ is a dimensionless LEC of order unity.  This term captures
the non-commutation of a spin with rotational fluid motion, for which the
term spin-orbit coupling  has become standard in the context of heavy-ion
physics. The term has no suppression by the UV cutoff, and is therefore
the same for both the charm and bottom quark.  If the fireball has a net
angular momentum, then the vorticity vectors in different parts of the
fireball must sum up to a non-vanishing value.  So such a coupling of the
quark spin to the local vorticity can give rise to net spin polarization.

It is interesting to ask what effect higher dimensional terms will have.
They are suppressed by powers of $k/\OLambda$ where $\Vc k$ is the
momentum of the phonon in the HQRF.  Since there is another vector,
namely the fluid velocity $\Vc u$, it is possible to construct an axial
vector $\Vc q=\Vc u\times\Vc k$. As a result, the simplest spin-dependant
coupling to a phonon is $\bsigma\cdot\Vc q$. This dimension 5 piece is
\beq
 L_5 = \frac{c_5'}\OLambda \epsilon^{\mu\nu\lambda\rho}
    v_\mu\,\bilin{\gamma_5\gamma_\nu}u_\lambda\partial_\rho\phi
   \heq \frac{c_5'\gamma}\OLambda\,\bilin{\bsigma_i}\,
     \epsilon^{ijk}\Vc u_j\Nabla_k\phi.
\eeq{hqsdcoup}
Since the direction of $\Vc k$ is random, averaging the resulting
polarization over all possible directions of $\Vc k$ gives a
vanishing result.  This argument also implies that corrections to
$L_w$ can only come from terms which involve $|\Vc k|^2$ multiplying
$\bsigma\cdot\Vc w$. These are dimension-6 terms, and are suppressed
by $k^2/\OLambda^2$. So, the universality between bottom and charm
polarization due to the coupling to vorticity is expected to be
numerically accurate.

\section{Application to heavy-ion collisions}

We will work with PDG values for the heavy quark masses
\cite{ParticleDataGroup:2020ssz}, namely $m=4.18\pm0.03$ GeV for
bottom quarks and $m=1.280\pm0.025$ GeV for charm. The best estimates
of the possible energy density in the early stages of a heavy-ion
collision currently come from Bayesian fits \cite{Bernhard:2016tnd,
JETSCAPE:2020mzn}. We can take the temperature reached at LHC energies
to be about 0.5 GeV, and in the top beam energy at RHIC to be about
0.3 GeV.  This gives almost an order of magnitude separation between
the bottom quark mass and both initial fireball temperatures, but
significantly less for the charm quark. So HQET in matter could be
quantitatively predictive for the bottom, but marginal for the charm.
Since we would like to satisfy the double inequality $m<\Lambda\ll T$,
a safe choice is to take $\Lambda\approx\sqrt{mT}$. This can be satisfied
with $\Lambda=2$ GeV for the bottom. At this scale the QCD coupling,
$\alphas\approx0.3$, which could allow computation of the LECs in the
weak coupling expansion. Pushing the UV cutoff to smaller values in order
to accommodate charm will cause us to lose this advantage. However,
as discussed previously, it is possible in this case to do a lattice
computation to match the LECs of HQET in matter.

For the construction of HQSD we need to consider one more time scale,
the fireball lifetime. For both top RHIC energy and LHC, one can take the
inverse of the expansion rate, $\M$, to be the fireball lifetime, \ie,
${\cal O}(10)$ fm, giving $\M\approx20$ MeV. Since we are using this
as an order of magnitude, it covers the range between 6 fm and 20 fm,
which is wide enough to accommodate the fireball lifetime in this range
of energies. With the values of initial $T$ already quoted, this implies
that the choice $\OLambda=50$--100 MeV is workable. It is possible that
the inverse of the expansion rate at an early stage of the fireball is
different from the lifetime. However, unless the inverse rate is as low
as 1--2 fm, this choice $\OLambda=100$ MeV still works at early times.

In this section we will concentrate on the polarization of high-$\pt$
heavy quarks at central rapidity. When $\beta_\CM\gamma_\CM=\pt/m$ is
large, the speed of the heavy-quark is larger than the speed of sound
in the fireball. As a result, the radial expansion of the fireball may
be neglected, and the heavy quark taken to leave the fireball at a time
which is no more than the nuclear radius $R_A\simeq 6$ fm. Conditions
in the fireball, such as the temperature and the vorticity will evolve
during this time. The values we use below should be considered to be
path-averaged. Due to the uncertainties in initial values of various
parameters we do not perform an actual averaging in this first estimate.

During the time that the fast heavy quark spends in the medium,
the coupling between its spin and thermal colour magnetic gluons, see
\eqn{hamint}, will cause the spin to relax. We present only an order of
magnitude prediction for the spin relaxation rate here, leaving a detailed
calculation to a later paper. Thermal fluctuations of the gluon field can
give rise to field strengths $gB\simeq T^2$. Their interaction with the
spin will relax in a typical time of order $m/T^2$, taking $c_2=1$. For
$T=0.5$ GeV, this leads to bottom quark relaxation time of a little more
than 3 fm, and the relaxation time approaches 6 fm for $T=0.4$ GeV.

The relaxation of the spin has two consequences at the LHC. Firstly, at
times of about 3 fm or so, one can use a thermal density matrix to a good
approximation. Second, because of the rapid decrease of the relaxation
rate as the temperature drops, the spin degree of freedom freezes out
and becomes a thermal relic pretty early. For fast heavy quarks, then
one can use a thermal density matrix with with a temperature of about
0.4--0.5 GeV after about 3 fm.

\bef[tb]
\begin{center}
 \includegraphics[scale=0.5]{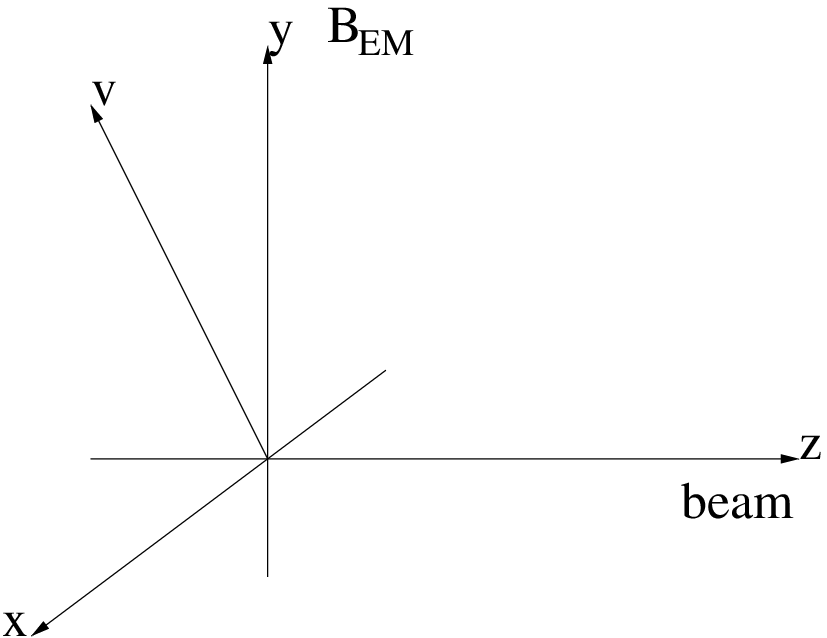}
 \includegraphics[scale=0.5]{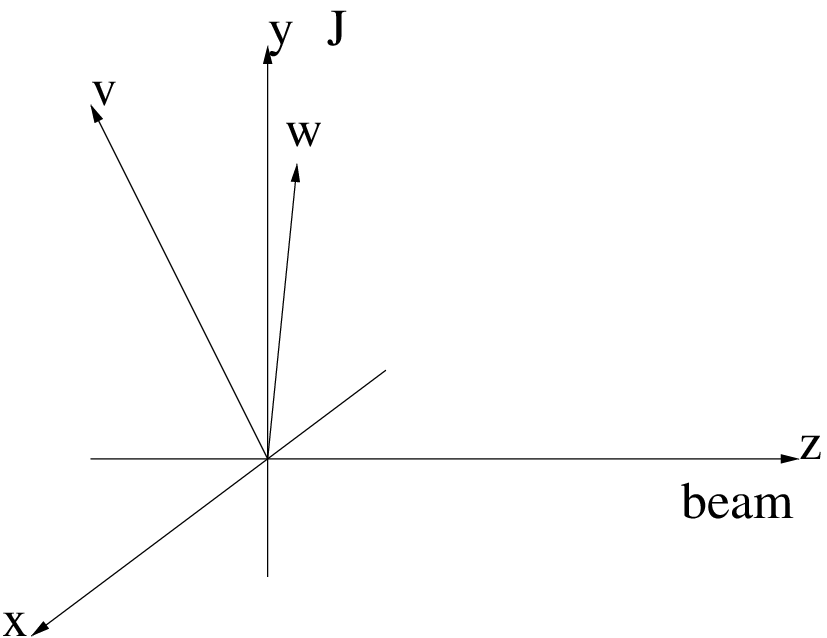}
\end{center}
\caption{In the CM frame of the colliding ions, the axes are oriented so
 that the beam is along the z-axis, and we align the x-axis so that
 the centers of the two colliding nuclei initially move along the
 xz plane. The initial magnetic field $B_\EM$ is oriented in the $y$
 direction, as is the net angular momentum, $\Vc J$ of the fireball. As
 shown in the figure on the left, the heavy quark velocity, $\Vc v$ is
 orthogonal to the beam (since we assume it is at central rapidity) and
 at an angle with respect to $B_\EM$. On the right we show a possible
 direction of the vorticity $\Vc w$ at the position of the heavy
 quark. Although $\Vc w$ may be oriented differently in different fluid
 elements, the heavy quark encounters a non-zero mean of $w$ along its
 path if $\Vc J$ is non-vanishing.}
\eef{frame}

All the heavy-quark spin Hamiltonians that we will need to deal with
are of the form $H=a+b\bsigma\cdot{\boldsymbol{\cal A}}$ where $a$
and $b$ are scalars, and rotational symmetry is broken by the axial
vector $\boldsymbol{\cal A}$. In the leading order in the EFT we need
only the single heavy-quark sector of Fock space. As a result, it is
sufficient to consider the two dimensional Hilbert space spanning the
spin states of the quark, and the treatment is an elementary exercise in
statistical mechanics. Using a $2\times2$ density matrix of quark spin
$\rho=\exp(-H/T)$, with $T$ the relic temperature, the polarization in
a direction specified by the unit 3-vector $\Vc n$ is given by $P={\rm
Tr\/}(\bsigma\cdot\Vc n\rho)/z$ where $z= {\rm Tr\/}\rho$. Quantizing
along $\Vc n$, one finds that
\beq
 P = \frac{{\rm e}^{\Delta E/T} - {\rm e}^{-\Delta E/T}}{
	 {\rm e}^{\Delta E/T} + {\rm e}^{-\Delta E/T}} 
   = \tanh\left(\frac{\Delta E}T\right)\qquad
 {\rm where}\qquad \Delta E=2b\Vc n\cdot\boldsymbol{\cal A}.
\eeq{pol}
In heavy-ion collisions, we will take the Hamiltonian in the CM frame
of the fireball, where the axes are chosen as shown in \fgn{frame}.
Since the initial magnetic field, $\Vc B_\EM$, as well as the net angular
momentum, $\Vc J$, in non-central collisions is oriented in the $y$
direction, this is the polarization predicted along this axis.

In the early stages of the fireball, it is estimated that
$eB_\EM=\zeta\tco^2$, where $\zeta={\cal O}(1)$ in the CM frame of the
colliding system (see \cite{Skokov:2009qp, Gursoy:2014aka}, for example),
and points in the $y$ direction. There is also an electric field in the
$x$ direction of similar magnitude; the numerical differences between
the magnitudes of $B_\EM$ and $E_\EM$ can be captured in different values
of $\zeta$, with both of order unity. We can write
\beq
  \Delta E = \frac{c\gamma_\CM\tco^2}{4m}, \qquad {\rm where}\qquad
  c=c_2q\zeta,
\eeq{bene} 
where $q=-1/3$ is the the charge of the bottom quark and $c_2=1+{\cal
O}(\alpha_\S)$. The biggest uncertainty is in the value of $\zeta$.
This large magnetic field will certainly polarize heavy quarks, but
the subsequent relaxation will wipe out the initial polarization. Any
observable polarization must be due to the magnetic field available at the
time that the spin freezes out. If this happens at a time of around 3--4
fm, then we need estimates of $\zeta$ at that time. It has been estimated
that after the first fm of lifetime, the EM fields decay extremely slowly
\cite{Gursoy:2014aka}, and one should be able to use $\zeta=0.01$. Then,
with $c={\cal O}(0.01)$, $\Delta E$ is of the order of a few tens of
KeV for $\pt$ ranging from 5--50 GeV. For such a small value of $\Delta
E$, one expects $P=\Delta E/T$ to good precision, which implies that the
polarization is far below the level of a percent. We show detailed results
in \fgn{initmag} with the choice $c=0.01$, for two values of $T$. The
bands of uncertainty shown come from uncertainties in $m$ and $\tco$.

\bef[bt]
 \begin{center}
	 \includegraphics[scale=0.7]{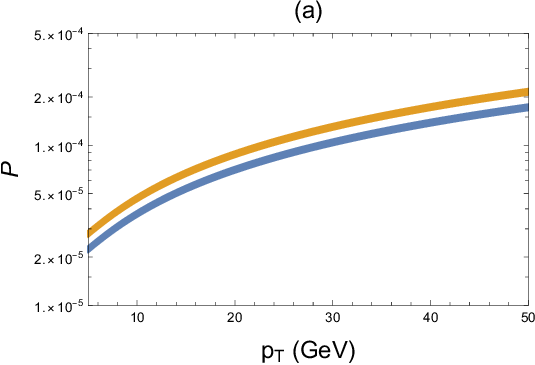}
	 \includegraphics[scale=0.7]{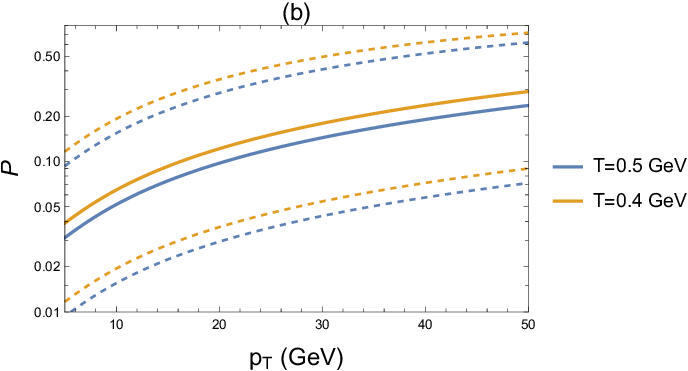}
 \end{center}
\caption{The $\pt$ dependence of bottom quark polarization, $P$, due to
 (a) the remnant magnetic field, $\Vc B_\EM$, at time of about 3 fm,
 and (b) the vorticity $\Vc w$.  In (a) have taken $|\Vc B_\EM| =
 \zeta\tco^2$, and used $c_2q\zeta=0.01$; the result is unlikely to
 be observable. The bands include the uncertainty in the bottom quark
 mass and $\tco$, which has been taken as $155\pm3$ MeV. In (b) we have
 chosen $2c_4|\Vc w_y|=10$ MeV (full lines).  The uncertainty comes
 from changing this combination from 3 to 30 MeV (dashed lines).}
\eef{initmag}

It is useful to note a difference between thermal averaging and averaging
over events. The result of \eqn{pol} is obtained by thermal averaging.
For the polarization due to colour fields produced by the Weibel
instability, one has $\Delta E\propto\Vc B_y$. This gives a spin
polarization $P=\Delta E(\Vc B_y)/T$ in one event. Since polarization
measurements average over events, the observed polarization will be
\beq
  P = \frac1{N_{evt}} \left(\frac{c_2 g}{2mT}\right)
	    \sum_{i=1}^{N_{evt}} \Vc B_y = 0,
\eeq{weibel}
where the number of events is $N_{evt}$. The sum vanishes because from
one event to another the strong field $\Vc B$ has independent orientations
and magnitudes. In the language of statistical mechanics $\Vc B$ is
a quenched random variable, and its effects could become measurable only
if event-by-event spin measurements were possible. Another example of 
quenched randomness could occur if the magnitude of $\Vc B_\EM$ varied
strongly from one event to another. In that case an averaging over
$\zeta$ would have to be performed. The result would still be non-zero
because $\Vc B_\EM$ always points in the same direction. However, this is
an academic discussion, since the polarization due to EM fields is
predicted to be so small.

On the other hand, for the coupling of the spin to the vorticity we have
\beq
 \Delta E = 2c_4\gamma_\CM\Vc w_y,
\eeq{vort}
where we can take $c_4\approx1$, and the polarization is measured with
respect to the $y$ direction, as before. Typical values of $\Vc w_y$
at the LHC are expected to be \cite{Huang:2020dtn} 10 MeV or so,
but can depend significantly on impact parameter and time. Given
this, we assume a central value of $2c_4\Vc w_y=10$ MeV, and use a
generous margin of uncertainty, 3--30 MeV. The results are shown in
\fgn{initmag}. The polarization of the heavy quark is predicted to
be large, and would definitely be observable. The value of $\Delta E$
is expected to be similar for bottom and charm quarks, and differences
between their polarizations, if any, would be evidence for a change in
their spin-freezeout temperature, $T$, which enters through \eqn{pol}.

\section{Conclusions}

In this paper we investigated heavy quark spin dynamics in strongly
interacting matter using heavy quark effective field theories
(EFT). These offer a controlled but extreme simplification of the
problem by reducing the Fock space of heavy quarks to the Hilbert space
of a single heavy quark in the leading order of the EFT. In particular,
for the polarization, this means that we need only to work within the
two dimensional space of a single heavy quark spin. We used two EFTs to
investigate the coupling of a heavy quark spin to the fireball at two
different energy and momentum scales.

The first is the analogue of HQET in matter. The HQET in vacuum is
widely used and tested in heavy-hadron phenomenology. The operator
product expansion that is used to write down the terms changes mildly
in matter as long as the UV cutoff of HQET maintains the hierarchy
$m>\Lambda\gg\tco$ (since $\tco\approx\lqcd$). We discussed
that this may be a reasonable approximation for bottom quarks, but is
more problematic for charm. We also discussed how the HQET for charm
would differ, and how its LECs can be matched to lattice QCD computations.

The second EFT is HQSD, and is valid at hydrodynamic scales, \ie, below a
UV cutoff $T>\OLambda \gg\M$, where $\M$ is the inverse of a typical
size (or time) scale of the fireball. This theory would be equally valid
for charm and bottom quarks. We gave a bottom-up construction of HQSD
which yielded a mass dimension-4 coupling between heavy quark spin and
the local vorticity of the fluid. This term does not have an explicit
dependence on the mass of the heavy quark, and should impart equal
polarization to the charm and bottom. We checked that the contribution of
dimension-5 terms to the polarization would vanish, so any corrections to
the universality of charm and bottom would come at best at dimension-6,
and hence is expected to be small.

We argued that HQET in matter leads to a thermal relaxation of the
heavy quark spin even when the quark is fast enough that its momentum
is not thermalized. An estimate of the relaxation time showed that a
bottom quark spin could be in thermal equilibrium for a relatively short
time. It is interesting to note that, depending on the value of the LEC,
the relaxation time of a charm quark could be shorter. As a result,
the spin of a fast charm quark might stay in thermal equilibrium for
longer. We showed that a consequence of the long time scale of bottom
quark spin relaxation is that the polarization due to the remnant EM field
would be small, and below the threshold of measurement. The observable
spin polarization of the heavy quarks would then be due to the fluid
vorticity, and is expected to be large (see \fgn{initmag}).

We also discussed two semi-quantitative predictions. The most basic of
these is that the spin polarization increases as $\gamma_\CM$ as predicted
by \eqn{pol}. This dependence on $\gamma_\CM$ indicates that the heavy
quark spin is coupled to a material property that is consistent in the
CM frame of the collision, \ie, in the rest frame of the fireball, and
is independent of heavy quark $\pt$. As one can see from the formalism
of Appendix A, this is a kinematic argument that lies at the base of
heavy quark EFTs, and constitutes a test of this approach.

The second prediction is our fairly robust estimate that the coupling
of the fluid vorticity to the heavy quark produces a polarization of
a few tens of percent. There is a single parameter that underlies the
prediction, which is the combination $2c_4\Vc w_y$ of \eqn{vort}. A test
is that the heavy quark polarization derived from spin alignment of
vector mesons such as $D^*$ and $B^*$ should agree with those derived
from the spin polarization of bottom and charm baryons. Furthermore,
by appropriate binning in collision centrality, one may further check
that the centrality dependence of this parameter is the same as that
expected of the angular momentum of the fireball.

Before concluding we would like to point out some directions which have
not been discussed in this work. We have not explicitly discussed the
role of three space-time symmetries which are interesting, namely C, P, and
T. In this work we have taken the thermal matter part of the action to be
symmetric in each. However, at finite chemical potential CP is violated,
and this will effect HQET and HQSD. The violation of parity in the fireball
has also attracted much attention. This symmetry breaking certainly would
have consequences for both HQET and HQSD. Both effects can introduce new
spin alignment terms, and will be of interest.  We leave such a study
for a future paper. Two aspects of hydrodynamics we have not remarked on
are of spin hydrodynamics and MHD. Both can affect observables. However,
their treatment requires detailed computation of flows, and lies beyond
the scope of this paper. We plan to investigate them separately.

I would like to thank Saumen Datta, Subrata Pal, and Rishi Sharma for
discussions.

\appendix

\section{Frames, matter, and HQET}
\label{sec:frames}

Since the heavy quark 4-velocity, $v$, is a constant timelike vector in
EFTs of heavy quarks, we have a special inertial frame, the heavy quark
rest frame (HQRF), in which one can write $v=(1,\boldsymbol 0)$. Many of
the Lorentz invariant arguments that we use simplify in the HQRF. However,
one can make their Lorentz invariance explicit by defining two projection
operators
\beq
 \PP_1^{\mu\nu} = v^\mu v^\nu, \qquad{\rm and}\qquad
 \PP_2^{\mu\nu} = g^{\mu\nu}-\PP_1^{\mu\nu}.
\eeq{proj}
We use the mostly negative metric $g={\rm diag\/}(1,-1,-1,-1)$.  One can
use the above partition of the metric tensor to decompose any 4-vector,
$a$, into parts parallel and orthogonal to $v$, namely the timelike piece
$a_\nu \PP_1^{\mu\nu} = (a\cdot v) v^\mu = a^0 v^\mu$, and the spacelike
part $a_\nu \PP_2^{\mu\nu} = a_\mu - a^0 v^\mu$.  This implies that
the decomposition $a=(a^0, \Vc a)$ in the HQRF is Lorentz invariant.
The decomposition of a derivative operator in the HQRF is therefore
also invariant

\beq
 \partial \heq \left(\frac{\partial}{\partial t},\Nabla\right), 
  \qquad{\rm where}\qquad
 \frac{\partial}{\partial t} = v\cdot\partial, \qquad{\rm and}\qquad
 \Nabla = v(v\cdot\partial) - \partial.
\eeq{decom}
Furthermore, every scalar product of two 4-vectors, $a$ and $b$, can
be invariantly decomposed into two pieces, namely $\PP_1^{\mu\nu}a_\mu
b_\nu = a^0b^0$ and $\PP_2^{\mu\nu}a_\mu b_\nu = \Vc a\cdot\Vc b$.

The heavy quark subspace $Q=P_+\psi$ (where $\psi$ is the full Dirac
spinor) is easily seen to correspond to the ``large components'' of the
spinor. The antiquark sits in the complementary subspace $q=P_-\psi$
of the ``small components''. These are down by a factor $1/M$.  As a
result, physics insights come easily by writing heavy-quark bilinears
in the HQRF in the representation of the Dirac matrices in which
\beq
 \gamma_0=\begin{pmatrix} {\mathbb I} & 0\\ 0 & -{\mathbb I}\end{pmatrix},
	 \qquad{\rm so\ that}\qquad
 \gamma_5=\begin{pmatrix} 0 & {\mathbb I}\\ -{\mathbb I} & 0\end{pmatrix}.
\eeq{reps}
In this representation, we find
\beqa
\nonumber
 && \bilin{\gamma_5} = \bilin{\gamma_i} = \bilin{\gamma_5\gamma_0}
  = \bilin{\Sigma_{0i}} = 0, \\
 && \bilin{\gamma_0}=\bilin{}, \quad 
    \bilin{\gamma_5\gamma_i}=\bilin{\sigma_i}, \quad 
    \bilin{\Sigma_{ij}}=\bilin{\sigma_i} = \epsilon_{ijk}\bilin{\sigma_k},
\eeqa{diracs}
where $i,\;j$ are spatial indices and $\sigma_i$ are the Pauli matrices.
The bilinears which vanish are those with (block) off-diagonal structure
to the Dirac matrices, for example, $\gamma_5$ as shown in \eqn{reps}.
Of course, physics does not depend on the representation of the Dirac
matrices, as one can see from checking that these statements simply
reflect whether or not $P_+\Gamma P_+$ vanishes.

The identities in \eqn{diracs} may be surprising at first,
because they seem to imply that $\overline\psi \slashed\partial \psi
\heq\bilin{ \gamma_0\partial_t} = \bilin{\partial t}$. This is clearly
an over-simplified form of the kinetic term for heavy quarks. The
explanation is that $\bilin{\slashed\Nabla}$ vanishes only to leading
order in $M$, so that the integration over diagrams containing virtual
anti-quark propagators gives rise to the remainder of the kinetic term,
namely $\bilin{\nabla^2}/(2M)$. Note that this is suppressed by one
power of the UV cutoff $M$ (see \cite{Neubert:1996wg} for a detailed
pedagogical discussion). Similarly, the other bilinears in \eqn{diracs}
get quantum corrections at higher orders in $1/M$.

In the application to heavy-ion collisions, we will need to
use the CM frame of the fireball. We use the notation $u_\CM$
to denote the 4-velocity of the CM in any inertial frame. In
the CM frame $u_\CM=(1,\boldsymbol 0)$, so that $\slashed
u_\CM=\gamma_0$. The boost between HQRF and the CM frame is $v\cdot
u_\CM=\gamma_\CM$. If the heavy quark is produced at central rapidity
with transverse momentum $\pt$, then, since $v=\gamma_\CM(1,\Vc
v_\CM)$, one has $\beta_\CM\gamma_\CM=\pt/m$, which implies that
$\gamma_\CM=\sqrt{1+\pt^2/m^2}$.

For applications, it will also be useful to write the Lagrangians in
the CM frame rather than in the usual HQRF. Then, instead of directly
writing the bilinears as in \eqn{diracs}, it is simpler to use commutators
with $P_+$. For our purposes, the most useful of these is 
\beq
 [P_+,\Sigma_{\mu\nu}]F^{\mu\nu}=\gamma_\CM
     (\Vc v\cdot\Vc E\gamma_0+\bsigma\cdot\Vc B),
\eeq{intlag}
since it directly allows us to write spin interactions in the CM
frame. A straightforward way to see this is to write out the definition
of $\Sigma_\mu\nu$ in terms of the Dirac matrices, and in the CM frame,
write $\gamma_0=\slashed u_\CM$, and $\epsilon^{ijk} \gamma_i \gamma_j
= \gamma_5 \slashed u_\CM \gamma_k$, where $i$, $j$, $k$ are spatial
indices.

In Section \RNum4 we present a consideration of scales which shows that one
may not be able to construct a useful thermal HQET for charm quarks using
weak coupling theory. If physics at the scale of the charm quark has
to be isolated from that of thermal matter, then one has to push the UV
cutoff of HQET closer to the temperature of matter, and do the matching of
the LECs through a lattice computation.  In constructing this version of
thermal HQET, another ingredient is needed.  A material at thermodynamic
equilibrium requires the introduction of a velocity vector, $u^\mu$,
of the heat bath with respect to the observer.  This gives the new
projection operators in the direction of, and transverse to, $u$, namely
\beq
 \PPi_1^{\mu\nu} = u^\mu u^\nu, \qquad{\rm and}\qquad
 \PPi_2^{\mu\nu} = g^{\mu\nu}-\PPi_1^{\mu\nu}.
\eeq{projmat}
Since $u$ and $v$ cannot be taken to be equal in general, so the
pair of projectors $\PP_{1,2}$ are distinct from $\PPi_{1,2}$. We
introduce the invariant $v\cdot u=\gamma$. This means that in the
HQRF, since $v=(1, \boldsymbol 0)$, one may write $u=\gamma(1,{\bf
u})$ with $|{\bf u}|=\beta=1/\sqrt{1-\gamma^2}$. Similarly, in the
FLRF, since $u=(1, \boldsymbol 0)$, one can write $v = \gamma(1,
{\bf v})$, yielding $|{\bf v}|=\beta$. Of course, $\bf u$ and
$\bf v$ are oriented independently. As a result, one finds that
${\PP_a}^\mu_\nu{\PPi_b}^{\nu\lambda}\propto\gamma$ for $a,b=1$, 2.

$\PPi_{1,2}$ allow us to define in an invariant way
two invariant components of the field tensor $F^{\mu\nu}$, which are
\beq
 D^{\mu\nu} = \PPi_1^{\mu\lambda}\PPi_1^{\nu\rho}F_{\lambda\rho}, 
    \qquad{\rm and}\qquad
 H^{\mu\nu} = \PPi_2^{\mu\lambda}\PPi_2^{\nu\rho}F_{\lambda\rho}.
\eeq{handd}
The notation has been chosen to remind us that in the frame co-moving with
the material, \ie, the fluid's local rest frame (FLRF), $D$ corresponds
to the electric field in matter, and $H$ to the magnetic field. In other
words, in the FLRF, since $u=(1,\boldsymbol 0)$, the structure of the
two tensors correspond exactly to the electric and magnetic components of
$F$. So $D_{0i}=-D_{i0}$ are the only possible non-vanishing components
of the tensor $D$ and $H_{ij}=-H_{ji}$ the only possible non-vanishing
parts of $H$, where $i$ and $j$ are spatial indices.

All this is perfectly in agreement with our knowledge of electrodynamics,
and other gauge theories, in matter. Since there are two independent and
orthogonal field tensors in matter, the Lagrangian of the gauge field
in matter can be written using two independent coefficients, $\epsilon$
and $\mu$, as
\beq
 L_{EM} = \frac\epsilon2D^{\mu\nu}D_{\mu\nu} + \frac\mu2H^{\mu\nu}H_{\mu\nu}.
\eeq{lagem}
Cross terms do not exist since the two tensors are orthogonal projections.
The equations of motion from \eqn{lagem} are easy to write, and give
rise to a wave equation with the dispersion relation $\omega^2=c^2|{\bf
k}|^2$ with $c^2=\mu/\epsilon$.  It is also straightforward to use the
equations of motion to show that the solutions of the wave equation
come with three polarizations, two transverse and one longitudinal
\cite{Weldon:1982aq}. Using the microscopic theory of the medium, the
coefficients $\epsilon$ and $\mu$ can be computed, as they have been
for gauge theories.

The decomposition of \eqn{handd} has a consequence for HQET in matter.  
The dimension-5 term from vacuum HQET splits
\beq
 c_2 \bilin{\sigma_{\mu\nu} F^{\mu\nu}}\longrightarrow
 c_2' \bilin{\sigma_{\mu\nu} D^{\mu\nu}} + 
 c_2'' \bilin{\sigma_{\mu\nu} H^{\mu\nu}},
\eeq{hqetmedium}
with two different coefficients to the two gauge theory tensors. Each
of the coefficients, $c_2'$ and $c_2''$, may be computed in a thermal
weak coupling expansion. The ``magnetic field'' in the HQRF, \ie,
the spatial components of the gauge fields in that frame, will have
contributions from both $D$ and $H$. In a weak-coupling computation
for HQET in thermal matter, both the coefficients $c_2'$ and $c_2''$
can be written as $1+{\cal O}(\alphas)$.

\end{document}